# Magnetization and Magneto-resistance in Y(Ba$_{1-x}$Sr$_x$)$_2$Cu$_3$O$_{7-\delta}$ ($x$ = 0.00 - 0.50) superconductor

N. P. Liyanawaduge[1,2,3], Shiva Kumar Singh[2], Anuj Kumar[2], Rajveer Jha[2], B. S. B Karunarathne[3] and V.P. S. Awana[2*]

[1]Industrial Technology Institute, P.O Box 363, Baudhaloka Mawatha, Colombo-07, Sri Lanka

[2]Quantum Phenomena and Application Division, National Physical Laboratory (*CSIR*) New Delhi-110012, India

[3]Department of Physics, University of Peradeniya, Peradeniya (20400), Sri Lanka

Here we present the magnetic properties and upper critical field ($B_{C2}$) of polycrystalline Y(Ba$_{1-x}$Sr$_x$)$_2$Cu$_3$O$_{7-\delta}$ superconductors, which are being determined through detailed *ac*/*dc* susceptibility and resistivity under magnetic field (*RTH*) study. All the samples are synthesized through solid state reaction route. Reduction in Meissner fraction (the ratio of field cooled to zero field cooled magnetization) is observed with increasing Sr content, suggesting occurrence of flux pining in the doped samples. The *ac* susceptibility and resistivity measurements reveal improved grain couplings in Sr substituted samples. Consequently the inter-grain critical current density ($J_c$), upturn curvature near the $T_c$ in temperature dependence of upper critical field [$B_{C2}(T)$], and $B_{C2}$ are enhanced. Both $J_c$ and $B_{C2}$ increase in lower Sr substitution (up to $x$ = 0.10) samples followed by decrease in higher doping due to degradation in effective pining and grain coupling.

## 1. Introduction

Application of external mechanical pressure on layered cuprate superconductors have been an important tool in engineering their increased transition temperature ($T_c$). The increase of $T_c$ depends upon the chemical composition of the parent phase. On the other hand, another strategy, which leads to optimization of physical properties of a given superconducting system, is cation/anion substitution. Therefore reproduction of the structural changes and further increase in physical properties, which are induced by mechanical pressure, can also be achieved by appropriate chemical substitutions. An example is the partial substitution of Sr cations (r$_i$ = 1.31 Å) at Ba (r$_i$ = 1.47 Å) site in YBa$_2$Cu$_3$O$_{7-\delta}$ (Y-123) cuprate superconductor. But substitution of Ba by Sr in Y-123 doesn't increase additional carriers in CuO$_2$ planes as both are divalent.

However this causes decrease in $T_c$ and results in an unstable 123 phase. Almost in all cuprates, $T_c$ decreases, when Ba cations are replaced by Sr, except for $(La_{1-x}Ba_x)_2CuO_4$ [1].

Several explanations had been proposed to account this negative coefficient ($dT_c/dx$) occurrence in $Y(Ba_{1-x}Sr_x)_2Cu_3O_{7-\delta}$. F. Licci *et al.* [2-3] proposed a mechanism in terms of Bond Valance Sum (*BVS*) for Ba atoms, which decreases with Sr substitution due to reduction in ionic size. This prevents charge transfer from Cu(1) to Cu(2) site and subsequently results in decreased $T_c$. According to Ona *et.al* [4] the O content decreases slightly in Sr doped samples due to structural disturbances. Along with decrease in O content, reduction in cell volume causes decrease in $T_c$ [2-6]. Veal *et.al* [5] accounted this $T_c$ reduction to local structure distortions occurred due to Sr doping. On the other hand Zheng *et.al* [7] concluded by X-ray Absorption Fine Structure *(XAFS)* that local structure distortions are not responsible for reduction in $T_c$ and the identity of $CuO_2$-Y-$CuO_2$ sandwich structure in Sr doped and un-doped samples is more important. Although Sr substitution shows negative effect on $T_c$, it is interesting to investigate how it affects other physical parameters like $J_c$ and $B_{C2}$, which are useful in practical applications of these materials [8-9]. In this investigation we have explored major limiting factors with these parameters such as inter-grain weak links and poor flux pining. We have determined inter-grain critical current density $J_c$ and upper critical field through resistivity under magnetic field measurements. It is found that moderate Sr substitution causes flux pining with enhancement of grain couplings, which could enhance both $J_c$ and $B_{C2}$.

## 2. Experimental

Samples of $Y(Ba_{1-x}Sr_x)_2Cu_3O_{7-\delta}$ (with nominal composition of $x$ = 0.00, 0.025, 0.05, 0.075, 0.10, 0.25 & 0.50) are synthesized through conventional solid state reaction route. High purity (99.99 %) powders of $Y_2O_3$, $SrCO_3$, $BaCO_3$ and CuO with stoichiometric ratio are mixed in agate mortar. After initial grinding, calcination is done in alumina crucibles at 860°C in air for 36 h. Subsequent calcinations are done at 870°C, 890°C and 910°C temperatures for same duration with intermediate grinding. In each calcination cycle cooling is done slowly and samples are re-ground well before the next cycle. After final calcination, the samples are pressed into rectangular pellet form and sintered at 925°C for 48 h in air. Finally the samples are annealed with flowing oxygen at 750°C for 12 h, 600°C for 24 h and 450°C for 24 h and subsequently cooled in 6 h. The phase formation is determined through *X-ray* powder diffraction, using *Rigaku X-ray* diffractometer (Cu-$K_\alpha$). Rietveld analysis of all samples is performed using Fullprof program. The *ac/dc* susceptibility ($\chi$-*T*), Magnetization (*M-T*) and resistivity measurements with and without field *(RTH)* are done by Physical Properties Measurement System (*Quantum Design*-USA PPMS-14Tesla). The scanning electron microscopy (*SEM*) images of the samples are taken on *ZEISS* EVO MA-10 Scanning Electron Microscope.

## 3. Results and Discussion

The phase analysis and structural parameters are determined by Rietveld refinement of *XRD* patterns. Fitted and observed *XRD* patterns of the studied $Y(Ba_{1-x}Sr_x)_2Cu_3O_{7-\delta}$ [$x$ = 0.00, 0.025, 0.05, 0.075, 0.10, 0.25 and 0.50] samples are shown in Fig. 1. It confirms that all the samples are crystallized in nearly single phase orthorhombic *pmmm* space group. The lattice parameters and other fitted parameters are given in Table 1. In agreement to the earlier studies [2-6] lattice parameters decrease with increasing *x*, which is a clear indication that Sr substitutes at Ba site. The orthorhombic distortion defined as $S = 2\frac{(b-a)}{(b+a)}$ is also tabulated in Table 1. According to the calculated values, the orthorhombicity is slightly increased with *x* and is maximum for *x* = 0.25. For *x* = 0.50 it again decreases in agreement with earlier studies [2-6]. However, small impurity phases arises in *x* = 0.50 composition as increasing Sr content results in an unstable 123 phase. The *SEM* images of freshly fractured samples with *x* = 0.00, 0.05, 0.10 and 0.25 are given in Fig. 2 (a-d). It can be observed that doped samples consist of slightly smaller grains than pristine sample in agreement to ref. [6]. Also the doped samples show better surface texture with lower level of porosity than pristine sample. These observations are in consistent with improvement of grain couplings with Sr doping.

Variation of dimensionless *dc* volume susceptibility, in both Field Cooled (*FC*) and Zero Field Cooled (*ZFC*) conditions is shown in Fig. 3. Decrease in $T_c$ with increasing Sr content can be seen. There is a remarkable increase in separation between *FC* and *ZFC* along with decreasing *FC* signal. This variation is monotonic up to *x* = 0.10. The increase in separation of *FC* and *ZFC* signal or in other words reduction in Meissner fraction (ratio of field cooled to zero fielded cooled magnetization) is a clear indication of flux pining [10]. This may enhance the critical current and critical fields. For compositions with *x* > 0.10 the *FC* and *ZFC* signal separation is reducing. The saturated magnetization of *ZFC* signal is a measure of diamagnetic shielding current, which initially increases with Sr content and then decreases in higher doping levels.

*ac* magnetization of all these samples measured at different *ac* field amplitudes with zero bias *dc* field is given in Fig. 4 (a-f). The driven magnetic field is parallel to the long axis of the measured samples. It is well known that the real part of susceptibility consists of two transitions, which correspond to the flux removal from intra-grain and inter-grain regimes. In accord, imaginary part contains two peaks, which represent energy dissipation and *ac* losses due to the flux motion in intra-grain and inter-grain regions [11-16]. This suggests that the higher temperature peak in the imaginary part, whose amplitude is measure of grain size [14-16], arises due to the grains and lower temperature peak is due to the grain boundaries. Accordingly in Fig. 4 (a), pristine sample shows dual signal structure in both imaginary and real part measurements. But none of doped sample show clear two peaks structure in imaginary part though it is observed in real part. This might due to higher inter-grain couplings of doped samples, which could suppress the intra-grain component. Also in Sr doped samples, smaller grains are observed in *SEM* images [Fig 2 (a-d)] due to which, intra-grain component got suppressed [16]. In pristine sample one can notice that the position (temperature) of high temperature peak is almost insensitive to the amplitude of driven field (inset Fig. 4a), while the position of low temperature peak is sensitive to the field amplitude. A broadening in low temperature peak is also accompanied with this [16]. For the lower fields, it is difficult to distinguish intra-grain and inter-grain regions due to persisting strong inter-grain couplings. But at moderate fields separate inter-grain and intra-grain regimes and well distinguished inter-grain and intra-grain signals are observed [11-16]. Due to this reason the inter-grain peak in Fig. 4 (a) lies very close to the intra-grain peak at lower fields but it shifts to the low temperatures at higher fields.

In case of Sr doped samples, the shift of inter-grain peak from intra-grain peak decreases with Sr substitution from $x = 0.00$ to 0.10, [see Figs. 4 (b-d)]. For x = 0.25 and 0.50, the shift of inter-grain peak from intra-grain peak is further increased, [see Figs. 4 (e-f)]. This clearly shows that the gains coupling of doped Y-123 is improved with Sr content till $x = 0.10$, and is decreased for $x = 0.25$ and 0.50 samples. The same is also inferred from the *SEM* images. For more clarity in regards to the behavior of the inter-grain peak, the *ac* susceptibility under different driven fields of all these samples is given in Fig.5. It is more understandable here that the curves are not only shifted to higher temperature but also the transition width is reduced for the moderately doped samples, till $x = 0.10$. Clearly the fields (ac amplitudes), which are strong enough to separate inter and intra-grain interactions in pure sample, are not strong enough to separate these interactions in moderately (*x* up to 0.10) doped samples. This indicates that strong inter-grain couplings are induced in the system with the substitution of Sr. Somehow coupling peak again tends to shift to lower temperatures in the higher doping ($x > 0.10$) indicating decrease of inter-grain couplings.

Fig. 6 shows the real and imaginary components of ac susceptibility measured at 17 Oe and 333 Hz *ac* driven field. This figure gives an explicit explanation on relative variation of strength of inter-grain couplings of doped and un-doped samples. In real part, the pristine sample shows well separated two step transitions with positive slope at the onset part. In contrast to this, with Sr substitution the second transition shifts towards the higher temperatures and is over lapped with first transition. This results in a sharp single transition with negative slope at the onset part, indicating a strong coupling component [11-16]. However with further increase of dopant level ($x > 0.10$) two step transition is again appeared, indicating weakening of grain coupling. In accord, two peaks in imaginary part follow the same variation. These observations collectively reflect that moderately Sr doped samples posses enhanced grain coupling, which may lead to increase in critical current and critical fields. Increase of $J_c$ up to moderate Sr concentrations is also observed in earlier reports [8]. We calculated the temperature dependency of inter-grain critical current density $J_c(T)$ using the relation $J_c(T_P) = H_a/a$ employing in Bean's model [17]. Here $2a \times 2b$, where $a < b$, is cross section of sample, $H_a$ is amplitude of applied ac field and $J_c(T_P)$ is the inter-granular critical current density at $T_p$, where the temperature of the inter-grain peak. The calculated values are plotted with temperature $(T_P)$ in Fig. 7. $J_c$ is estimated only in neighborhood of $T_c$, where the shift of the inter-grain peak's extendibility towards lower temperatures is within the limit of driven field. It is seen that inter-grain current is improved for moderately Sr doped samples and excessive doping cause degradation of critical current. Both *dc* susceptibility (Fig. 3) and detailed *ac* magnetization (Figs. 4-6) results, clearly indicate that inter-grain coupling of pristine Y-123 is improved for moderately ($x$ up to 0.10) Ba site Sr doped samples.

Fig. 8 (a-e) depicts the normalized resistivity curves under magnetic field ($\rho_{nor}TH$) up to 13 T dc field, applied perpendicular to the current flow for the Y(Ba$_{1-x}$Sr$_x$)$_2$Cu$_3$O$_{7-\delta}$ [$x$ = 0.00, 0.05, 0.10, 0.25 and 0.50] samples. In zero field, the $T_c$ ($\rho$ = 0) of $x$ = 0.00 sample is around 90 K (Fig. 5a) and is decreased monotonically to around 80 K for $x$ = 0.50 (Fig. 5e) sample. The decrease in $T_c$ ($\rho$ = 0) is in agreement with the magnetization results (Figs. 3 and 4). *RTH* curves show [Figs. 8 (a-e)] basically two transitions, which is in agreement with magnetization and earlier reports [18-21]. It can be seen that the effect of magnetic field is weaker at onset part near normal state in comparison to the tail part. Also, off set of $T_c$ ($\rho$ = 0) is moved to lower temperatures with increasing field. This occurs near the onset part, where superconductivity persists only inside individual grains and superconducting fraction is quite small. On the other hand at the tail part, superconductivity persists not only in grains but also in grain boundaries, leading to higher superconducting fraction [22]. A long range superconducting state with zero resistance is achieved by means of a percolation like process that overcomes the weak links between grains [23]. The broadening of tail part occurs due to the anisotropic nature and disturbances in percolation path between grains (due to poor grain alignments) caused by applied field [24-25]. It is also noticeable that the rapid shift of the tail part towards lower temperatures is occurred at low fields (below 5 T) afterward almost small and uniform shift occurred at higher field. This can be explained as: in lower applied fields with increasing field, poorly oriented grains are becoming normal. But at higher fields, only favorably oriented grains contribute to superconductivity thus no further broadening. Also in higher magnetic fields, the field tends to penetrate individual grains and this cause broadening of onset part. As far as impact of Ba site Sr substitution is concerned, it can be noticed that in Sr doped samples there is a monotonic decrease in shift of tail part towards lower temperatures. This may be attributed to enhanced grain coupling in Sr doped samples in consistent with *ac/dc* susceptibility data. However in higher doping levels ($x$ = 0.25 and 0.50), increase in shifting of tail part may be due to slight weakening of grain couplings in complying with *ac* magnetization data.

The plots of temperature derivative of resistivity are shown in Fig. 9 (a-e). It is well known that temperature derivative of resistivity gives narrow intense maxima centered at $T_c$ and broad peak at low temperatures representing intra-grain and inter-grain regimes [18-21]. In all plots [Fig. 9 (a-e)] a single peak structure appears in zero field confirming that the grains maintain good percolation path between them. On the other hand two peaks are seen under applied field, corresponding to the both regimes, being separated from each other at different rates with increasing field. The peak corresponding to granular network is shifting towards lower temperatures more rapidly than the bulk peak. Simultaneously rapid broadening of coupling peak can be seen. The bulk peak is also getting diminished with increasing field as flux penetrates into the individual grains. It can also be seen that separation of coupling peak from bulk peak has decreased in Sr doped samples. We calculated the temperature difference ($\Delta T_{P2}$) between the positions of coupling peak at 13 T and at 0.1 T and plotted its variation with $x$ [see Fig. 10]. This shows the gap between two peaks reduces with increase of $x$ until $x \leq 0.10$ then slightly increases for $x > 0.10$. By taking all these into the account (*ac* susceptibility and resistivity data) it is reasonable to conclude that the moderate substitution of Sr at Ba site in Y-123 system causes enhancement of grains coupling.

The estimation of temperature dependency of resistive upper critical field [$B_{C2}(T)$] using mid point data (where the resistivity is half of its normal state value), is shown in inset of Fig. 10. It is seen that all the samples show concave curvature (upward curvature) near the $T_c$ followed by linear region which is in agreement with the early studies [24-26]. It is also noticeable that this upturn has increased in Sr doped samples than the pristine sample. Werthamer, Helfand, and Hohenberg (*WHH*) theory gives a solution for linearized Gor'kov equations for $H_{c2}$ for bulk weakly coupled type II superconductors, including effects of Pauli spin paramagnetism and spin orbit scattering [27]. Here we use simplified WHH equation to estimate $B_{C2}(T)$ without spin paramagnetism and spin orbit interaction given by

$$ln\frac{1}{t} = \psi\left(\frac{1}{2} + \frac{\bar{h}}{2t}\right) - \psi\left(\frac{1}{2}\right)$$

Where $t = T/T_c$, is the digamma function and $\bar{h}$ is given by

$$\bar{h} = \frac{4H_{c2}}{\pi^2 T_c(-dH_{c2}/dT)_{T=T_c}}$$

Using slope of linear portion of the experimental data and corresponding extrapolated $T_c$ values to $\mu_o H_{c2} = 0$ region, the $\mu_o H_{c2}(T)$ is fitted using simplified *WHH* model. The fitted data is shown in Fig. 11. In accordance with the pining behavior revealed in *dc* magnetization measurements and variation of inter-grain coupling strength, the estimated $B_{C2}$ value increases with Sr substitution followed by a small decrease. The calculated $B_{C2}(0)$ of pristine sample is around 66 T and with increasing Sr concentration maximum $B_{C2}(0)$ is found to be around 140 T for $x = 0.10$ composition. Several studies have been made earlier on single crystals of Y-123 and $B_{C2}(0)$ is calculated [28-31]. Large anisotropy has been observed for $B_{C2}(0)$ when field is applied parallel and perpendicular to $CuO_2$ planes [28-30]. Various approximations have been used and the calculated $B_{C2}(0)$ is varies from 56 T-120 T when field is applied perpendicular to $CuO_2$ planes [28-30]. However the same varies from 190 T-600 T when field is applied parallel to $CuO_2$ planes [28-30]. Our findings for the studied samples (66 T for pristine Y-123 and 140 T for $x = 0.10$ composition) seems reasonable in context of bulk polycrystalline samples.

Summarily, Sr substitution at Ba site results in an increase of flux pining property in Y-123. Effective pining initially increases with Sr content followed by a slight decrease. The grain coupling behavior in term of its strength follows same variation with Sr content. The inter- grain critical current density and resistive upper critical field increase with Sr doping due to the effective flux pining together with improved grain coupling nature of doped Y-123 samples.

## 4. Acknowledgement

This work is supported by *RTFDCS* fellowship programme conducted by Center for International Cooperation in Science (*CICS*). Authors are thankful to Prof. R.C Budhani, Director *NPL* and to Dr. (Mrs.) Ganga Radhakrishnan, Director *CICS* for their encouragement.

**Figure Caption and Table**

**Fig.1** Rietveld fitted powder *XRD* patterns of Y(Ba$_{1-x}$Sr$_x$)$_2$Cu$_3$O$_{7-\delta}$, $x$ = 0.00, 0.025, 0.05, 0.075, 0.10, 0.25 and 0.50 oxygen annealed samples.

**Fig.2** *SEM* images of Y(Ba$_{1-x}$Sr$_x$)$_2$Cu$_3$O$_{7-\delta}$; (a) $x$ = 0.00, (b) $x$ = 0.05, (c) $x$ = 0.10 and (d) $x$ = 0.25 samples at 5,000 magnification.

**Fig.3** Variation of dimensionless *dc* volume susceptibility with temperature of Y(Ba$_{1-x}$Sr$_x$)$_2$Cu$_3$O$_{7-\delta}$ $x$ = 0.00, 0.025, 0.05, 0.10, 0.25 and 0.50 samples.

**Fig.4** Real part and imaginary part of ac magnetization of Y(Ba$_{1-x}$Sr$_x$)$_2$Cu$_3$O$_{7-\delta}$ (a) $x$ = 0.00 (b) $x$ = 0.025 (c) $x$ = 0.05 (d) $x$ = 0.10 (e) $x$ = 0.25 (f) $x$ = 0.50 samples, taken at different ac driven field amplitudes with 333Hz constant frequency.

**Fig.5** The zoomed images of the coupling peak in ac susceptibility under different applied fields of Y(Ba$_{1-x}$Sr$_x$)$_2$Cu$_3$O$_{7-\delta}$, $x$ = 0.00, 0.025, 0.05, 0.10, 0.25 and 0.50 samples

**Fig.6** Real and imaginary part of ac susceptibility of Y(Ba$_{1-x}$Sr$_x$)$_2$Cu$_3$O$_{7-\delta}$, $x$ = 0.00, 0.025, 0.05, 0.10, 0.25 and 0.50 samples, performed at 17 Oe and 333Hz *ac* driven field.

**Fig.7** Variation of $J_c$ against $T_p$ of Y(Ba$_{1-x}$Sr$_x$)$_2$Cu$_3$O$_{7-\delta}$, $x$ = 0.00, 0.025, 0.05, 0.10, 0.25 and 0.50 samples. Solid lines are only guide to the eye.

**Fig.8** Temperature dependency on normalized resistivity taken under different dc fields ($\rho_{nor}TH$), applied perpendicular to the current flow of Y(Ba$_{1-x}$Sr$_x$)$_2$Cu$_3$O$_{7-\delta}$, (a) $x = 0.00$, (b) $x = 0.05$, (c) $x = 0.10$ (d) $x = 0.25$ (e) $x = 0.50$ samples.

**Fig.9** Temperature derivative of normalized resistivity of Y(Ba$_{1-x}$Sr$_x$)$_2$Cu$_3$O$_{7-\delta}$, (a) $x = 0.00$, (b) $x = 0.05$, (c) $x = 0.10$ (d) $x = 0.25$ (e) $x = 0.50$ samples, derived from resistivity data in Fig.8. The inset shows the zooming picture of coupling peak.

**Fig.10** The variation of temperature difference between positions of coupling peak at 13 T and 0.1 T with $x$. The solid line is guide for the eyes.

**Fig.11** Fittings of resistive upper critical field [$B_{C2}(T)$] using simplified WHH theory of Y(Ba$_{1-x}$Sr$_x$)$_2$Cu$_3$O$_{7-\delta}$ $x = 0.00$, 0.05 0.10, 0.25 and 0.50 samples. Inset shows the corresponding experimental temperature dependency on $B_{C2}(T)$ derived from resistive transitions.

**Table 1** Rietveld refined parameters and orthorhombic distortion of Y(Ba$_{1-x}$Sr$_x$)$_2$Cu$_3$O$_{7-\delta}$, oxygen annealed samples.

| Y(Ba$_{1-x}$Sr$_x$)$_2$Cu$_3$O$_{7-\delta}$ | a (A°) | b (A°) | c (A°) | R$_p$ | R$_{wp}$ | $\chi^2$ | $S = 2\dfrac{(b-a)}{b+a}$ |
|---|---|---|---|---|---|---|---|
| $x = 0.00$ | 3.822(9) | 3.886(7) | 11.681(9) | 5.34 | 7.02 | 3.68 | 0.0150 |
| $x = 0.025$ | 3.831(1) | 3.897(3) | 11.693(9) | 4.90 | 6.20 | 2.91 | 0.0161 |
| $x = 0.05$ | 3.826(8) | 3.883(7) | 11.684(4) | 4.96 | 6.25 | 2.50 | 0.0153 |
| $x = 0.075$ | 3.824(3) | 3.885(6) | 11.672(3) | 4.53 | 5.68 | 2.48 | 0.0163 |
| $x = 0.10$ | 3.813(7) | 3.884(3) | 11.654(5) | 5.02 | 6.31 | 3.07 | 0.0164 |
| $x = 0.25$ | 3.813(1) | 3.872(6) | 11.636(2) | 5.10 | 7.30 | 5.23 | 0.0171 |
| $x = 0.50$ | 3.792(7) | 3.851(7) | 11.595(3) | 6.20 | 9.41 | 10.7 | 0.0158 |

Fig.1

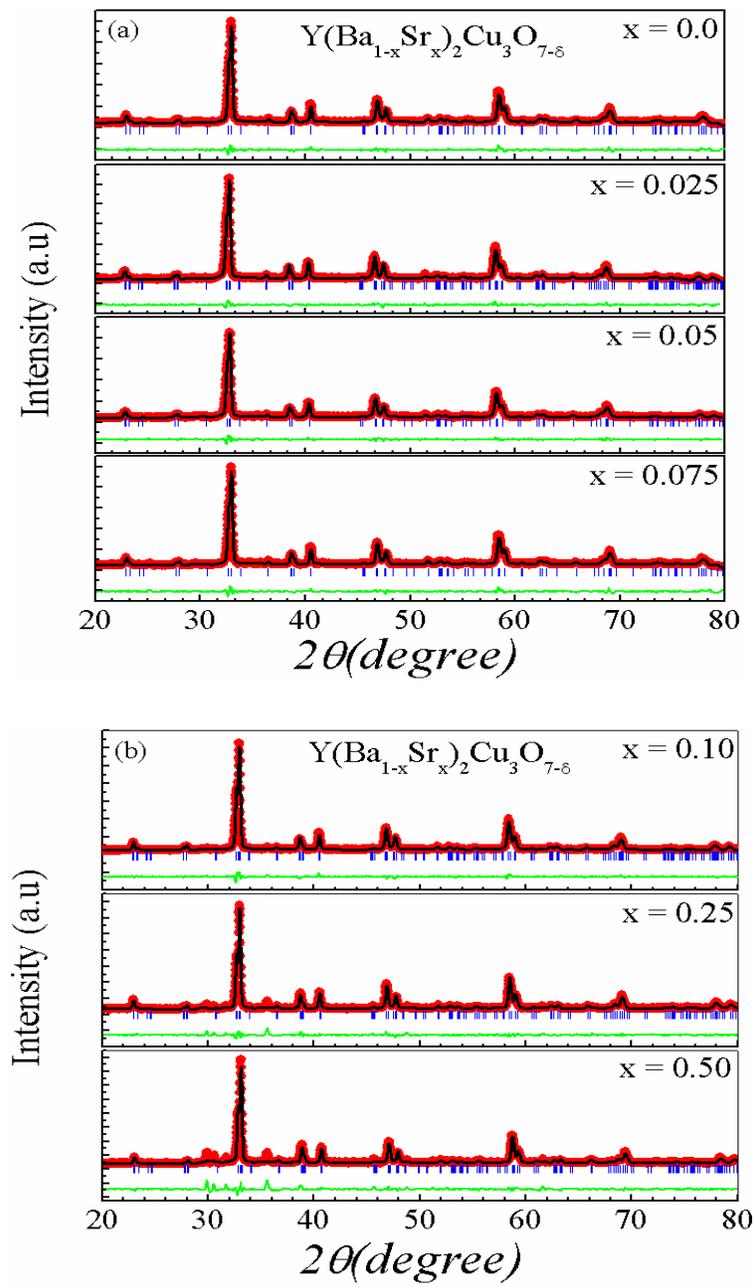

Fig.2

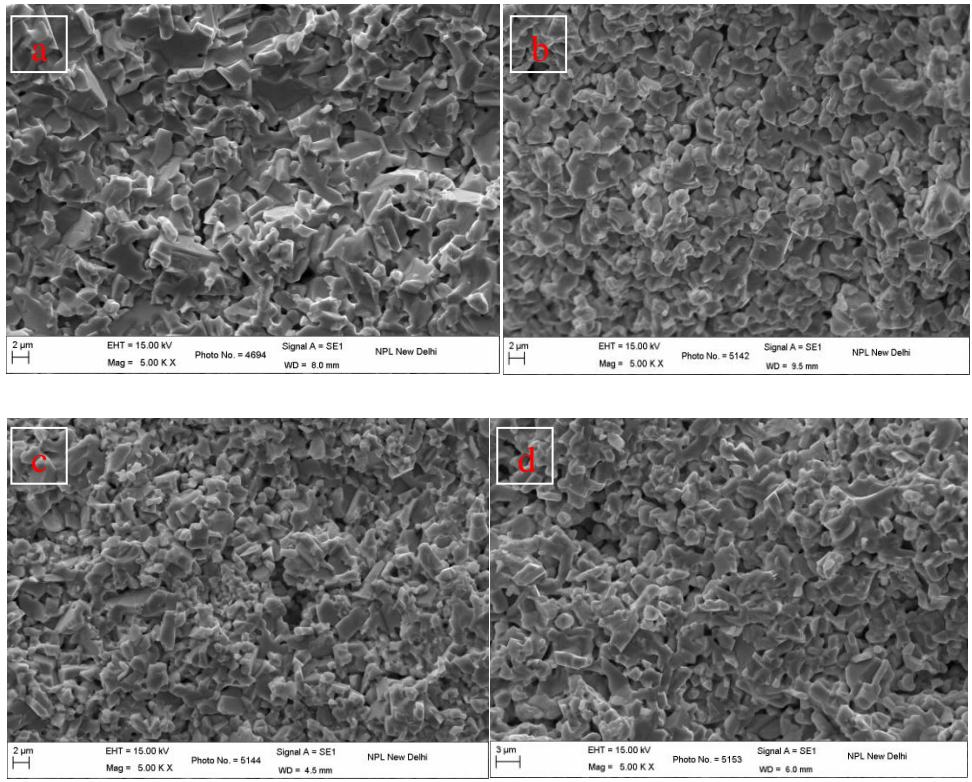

Fig.3

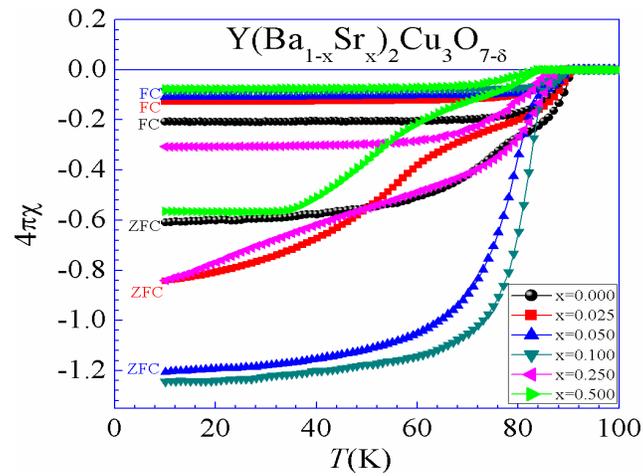

Fig. 4

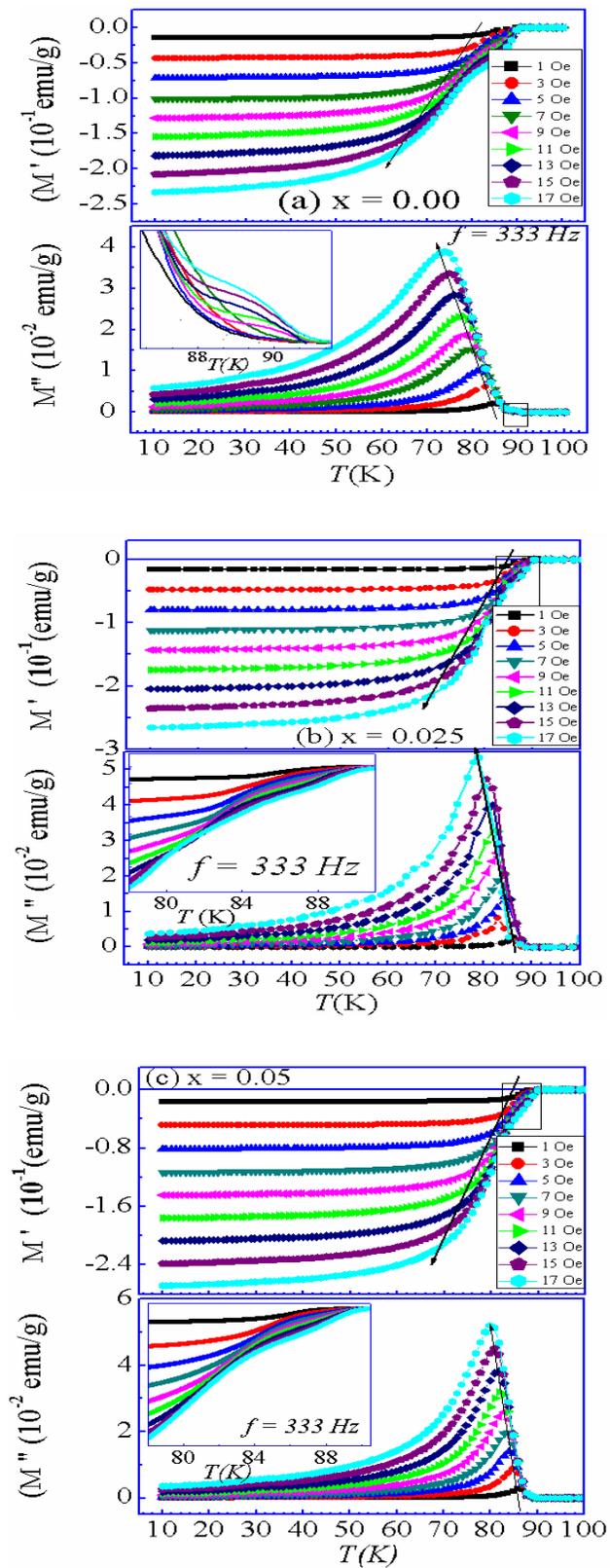

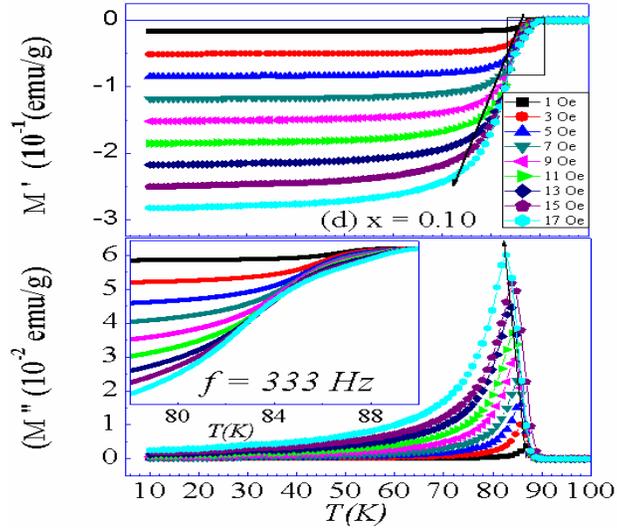

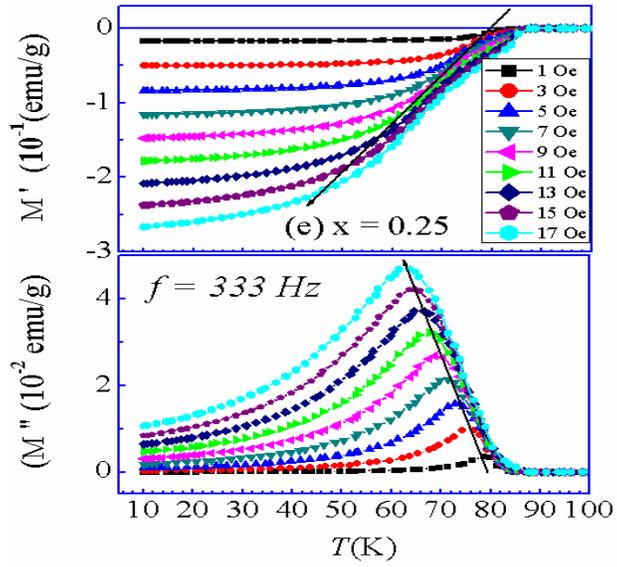

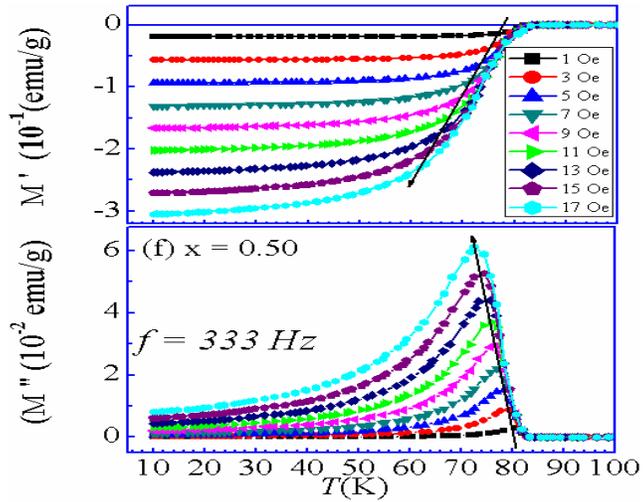

Fig.5

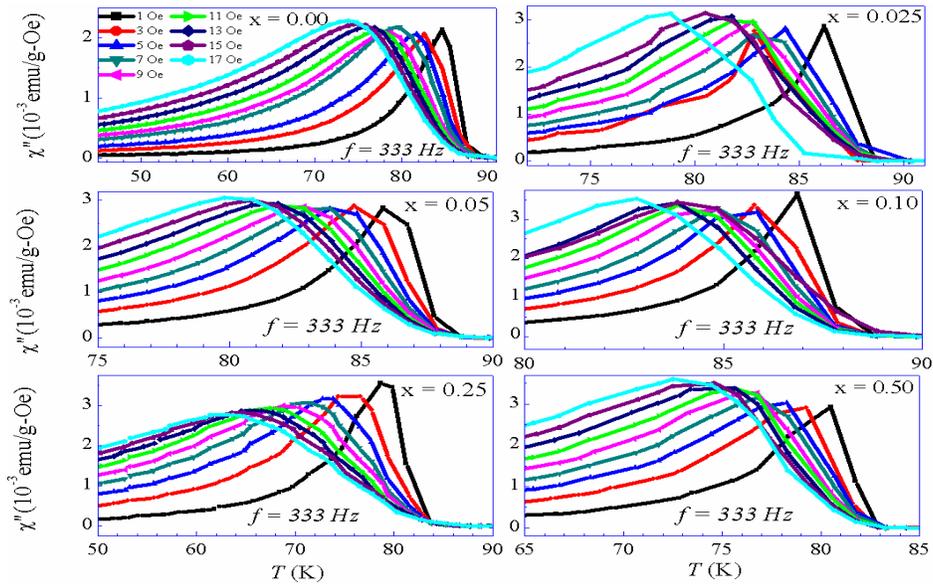

Fig.6

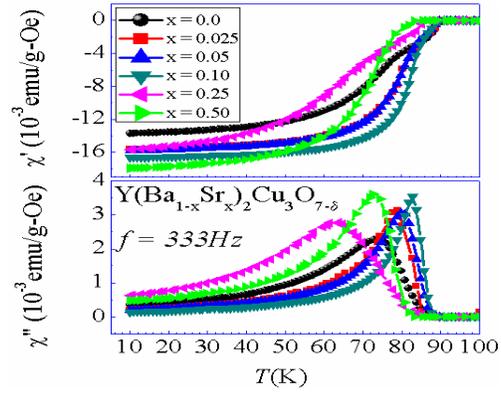

Fig.7

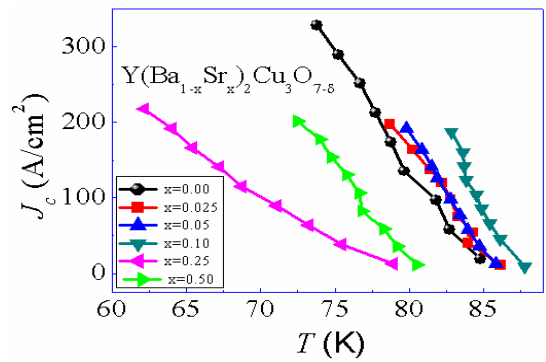

Fig.8

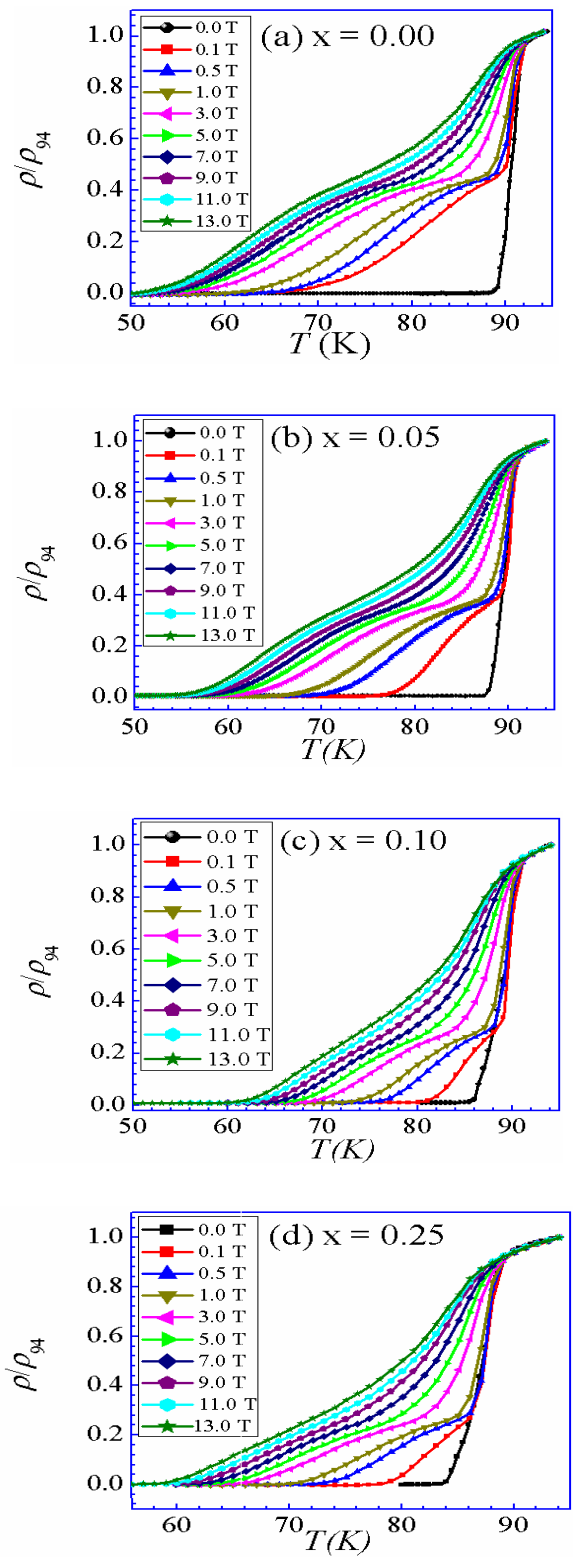

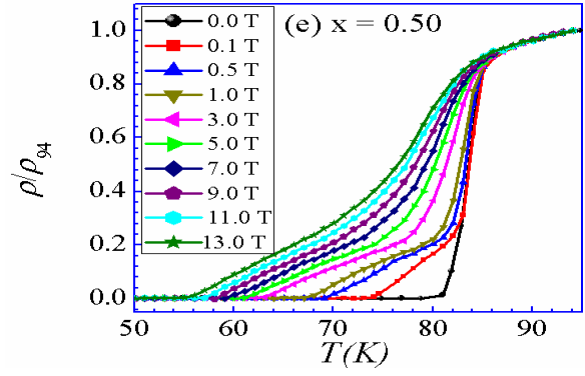

Fig.9

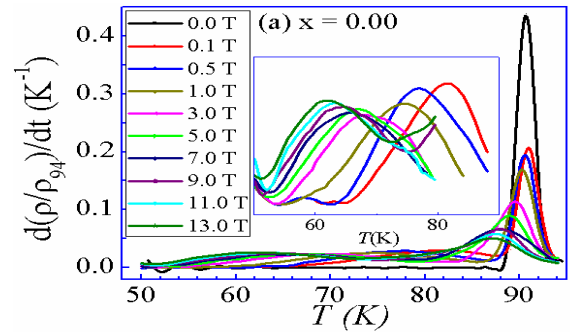

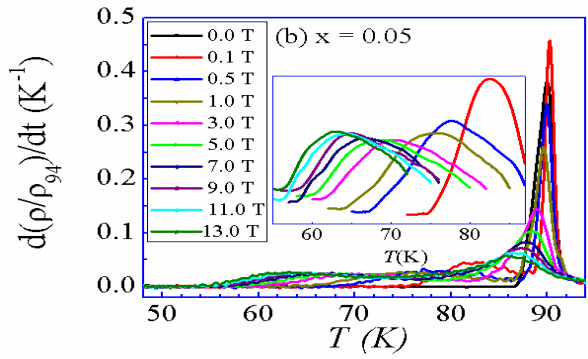

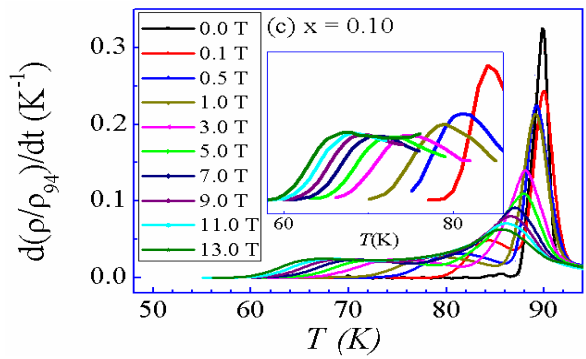

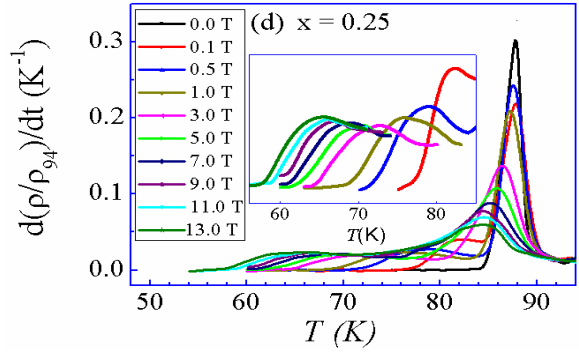

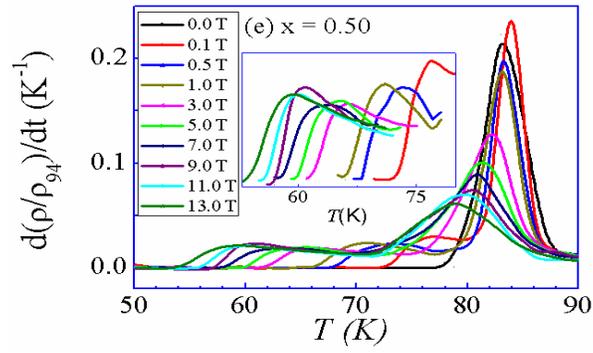

Fig.10 Fig.11

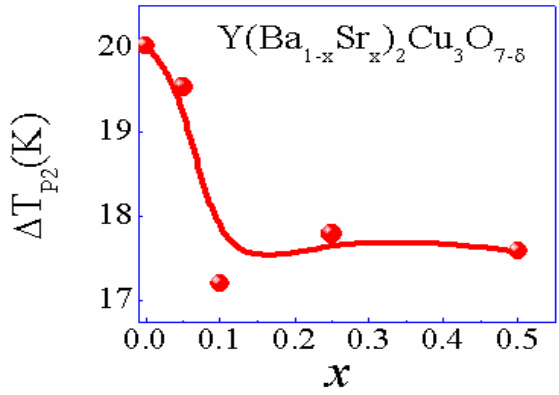

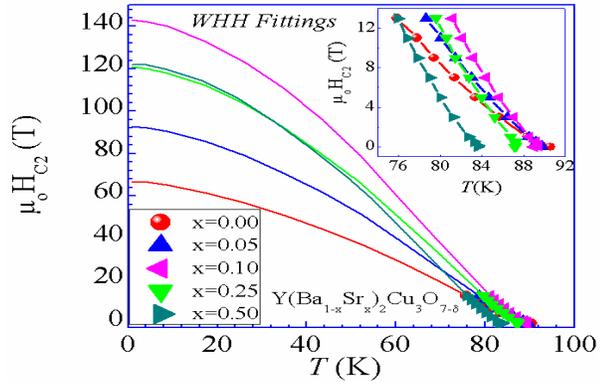

17